%% file: 2670main.tex
\documentclass{aa}
 \usepackage{graphicx}
 \usepackage{epstopdf}
\usepackage{txfonts}
\usepackage{lscape}

\def\square2{\vrule height 6pt width 6pt depth -0.5pt}

\def\rperi{$r_{\rm peri}$}
\def\Rcr{$R_{\rm cr}$}
\def\Re{$R_{\rm e}$}
\def\Ndwarfs{\ifmmode{N_{\rm\scriptscriptstyle D}}\else{$N_{\rm\scriptscriptstyle D}$}\fi}
\def\SigCR{\Sigma_{\rm\scriptscriptstyle CR}}
\def\lambdaCR{\lambda_{\rm\scriptscriptstyle CR}}
\def\sig1D{\sigma_{\rm\scriptscriptstyle 1D}}
\def\kms{\ifmmode{\rm{km\,s^{-1}}}\else{km\,s$^{-1}$}\fi}
\begin{document}


%\title{Harrassment Origin for Core-envelope Kinematic Decoupling in Dwarf Elliptical Galaxies?}
\title{Harassment Origin for Kinematic Substructures in Dwarf Elliptical Galaxies?}


\author{A. C. Gonz\'alez-Garc\'{\i}a \and J. A.
  L. Aguerri \and M. Balcells}
%\institute{Instituto de Astrof\'{\i}sica de Canarias, Via L\'actea s/n;
%  38200 La Laguna, Spain. \\cglez@iac.es, jalfonso@iac.es, balcells@iac.es}
\offprints{cglez@iac.es}
\institute{Instituto de Astrof\'{\i}sica de Canarias, Via L\'actea s/n;
  38200 La Laguna, Spain. \\cglez@iac.es, jalfonso@iac.es, balcells@iac.es}
%\email{cglez@ll.iac.es, jalfonso@ll.iac.es, balcells@ll.iac.es}

\date{Received <date> / Accepted <date>}
\authorrunning{Gonz\'alez-Garc\'{\i}a, Aguerri \and Balcells}
\titlerunning{Harassment Origin for KDC in dE's?}

\abstract{
We have run high resolution N-body models simulating the encounter of
a dwarf galaxy with a bright elliptical galaxy. The dwarf absorbs orbital angular momentum and shows
counter-rotating features in the external regions of the galaxy. To explain the core-envelope kinematic decoupling observed in some dwarf galaxies
in high-density environments requires nearly head-on collisions and very little dark matter bound to the dwarf. These kinematic structures appear under rather restrictive conditions. As a consequence, in a cluster like Virgo $\sim$ 1$\%$\ of dwarf galaxies may present counter-rotation formed by harassment.
%\mbremark{habra que cambiar esto si tomamos los nuevos calculos de probabilidades}

}


%% subject header.  Objects should be in the appropriate "individual"
%% headers (e.g. quasars: individual, stars: individual, etc.) with the
%% additional provision that the total number of headers, including each
%% individual object, not exceed six.  The \objectname{} macro, and its
%% alias \object{}, is used to mark each object.  The macro takes the object
%% name as its primary argument.  This name will appear in the paper
%% and serve as the link's anchor in the electronic edition if the name
%% is recognized by the data centers.  The macro also takes an optional
%% argument in parentheses in cases where the data center identification
%% differs from what is to be printed in the paper.

\maketitle


\section{Introduction}
\label{sec:Introduction}

%\mbremark{general: Unas veces tenemos $R_{ef}$, otras $R_e$.  Para asegurar notacion unificada, debemos usar un comando (lo he definido en e preambulo del fichero)}

Dwarf galaxies, the most numerous galaxy type in the Universe (Binggeli, Sandage, \& Tammann 1988; Ferguson \& Binggeli 1994; Gallagher \&
Wyse 1994), come in two groups: dwarf ellipticals (dE) and dwarf
irregulars (dI). These low mass systems have similar stellar
distributions, showing exponential surface brightness profiles with similar 
central surface brightness and scale lengths (Lin \&
Faber 1983). They also follow the same luminosity-metallicity relation
(Skillman et al.~1989; Richer \& McCall 1995). But, dEs and dIs differ in the gas
content, dIs being gas-rich galaxies while dEs are devoid of warm gas. 
%The similar stellar distributions between dE and dI suggest the presence of linked evolutionary scenarios between both classes of galaxies. 

Whether dIs and dEs are evolutionary linked, and what such links may be, is a matter of current debate.  Both internal and external processes have been proposed.  Among the internal processes, kinetic energy from supernova explosions can sweep the gas and turn dI galaxies into dEs (e.g. Dekel \& Silk 1986; De Young \& Gallagher 1990). However, such mechanism cannot explain recent evidence that the velocity gradients in most dE galaxies in Virgo cluster are generally different from those of dIs (van Zee, Skillman \& Haynes 2004).  

Environmental processes may be suspected given that dE galaxies are the most strongly clustered galaxy type (Ferguson \& Sandage 1989), hence   
encounters with other galaxies could be important in their evolution (Aguerri et al.\ 2004, 2005 and references therein).  van Zee, Skillman \& Haynes (2004) propose interaction-driven gas striping as the main mechanism for the transformation of dIs into dE galaxies.   Dwarf galaxies may also suffer transformations due to fast interactions with large cluster members and to their motion in the steep cluster potential, a process known as harassment (Moore et al.\ 1996; Mayer et al.\ 2001).  

Understanding the fossil records of evolutionary transformations in a high-density environment may help sorting out the relative importance of the above mentioned processes.  In a recent paper, de Rijcke et al.\ (2004) argue that fast interactions with giant galaxies may leave a signature in the rotation curve of dwarf galaxies, including core-envelope counterrotation.  In their deep spectroscopy of 15 dE galaxies, they report on two cases of dEs showing complex kinematic profiles.  FS373, a nucleated dE(2,N) in the NGC~3258 group, shows a 0.3 kpc ($\sim$0.2 \Re) kinematically-decoupled core (KDC), in corrotation with the main body, plus an outer velocity reversal at 2.2 kpc ($\sim$1.5 \Re).  FS76, a dE0 dwarf in the NGC~5044 group, shows a nuclear corrotating KDC in the inner $\sim$0.2 kpc (0.26 \Re) and an asymmetric rotation profile outside 1~\Re. 

On the basis of the large velocity dispersions of both groups, de Rijcke et al.\ (2004) argue that mergers are unlikely mechanisms for the origin of the complex kinematic profiles of the two dEs.  Instead, they propose that the velocity reversals could result from fast interactions with the dominant galaxies in their groups.  Their analytic estimates based on the impulse approximation show that sufficient angular momentum may be exchanged during a fast encounter to reverse the rotation of the dwarfs.  

Kinematic decoupled cores (KDC) in bright galaxies have been well
studied in the past two decades.  Most well known are counterrotating cores in elliptical galaxies (Bender 1990), which have been explained with mergers of unequal ellipticals (Balcells \& Quinn 1990), or with mergers of spiral galaxies, either with (Hernquist \& Barnes 1991) or without (Balcells \& Gonz\'alez 1998) dissipative gas dynamics.  Some kinematic reversals in elliptical nuclei might be the result of projection effects on triaxial ellipticals (Oosterloo, Balcells, \& Carter 1994).  Counterrotating bulges may originate in mergers of dwarf galaxies onto large disk galaxies (Aguerri, Balcells, \& Peletier 2001).  We note that counterrotation may appear in primordial collapses as well (Harsoula \& Voglis 1998).  

For dwarf galaxies, while mergers can in principle yield counterrotation, we concur with de Rijcke et al.\ (2004) that high group velocity dispersions make those highly improbable.  Other processes, studied for the cases of counterrotation in spiral galaxies, may apply to dwarfs.  These include: scattering of stars by a bar (Evans \& Collett 1994); the evolution of a polar ring in a triaxial halo (Tremaine \& Yu 2000); and even simple projection effects on triaxial or non-planar stellar disks.  

The harassment hypothesis put forward by de Rijcke et al.\ (2004) offers yet another mechanism for the origin of KDCs in dwarf galaxies.  In this paper we simulate, using high-resolution numerical models, the impulsive encounter of a dwarf galaxy with a giant elliptical, in order to perform direct measures of the true effect of the encounter on the rotation profile of the dwarf. Our aim is to test the validity of the analytical predictions of angular momentum exchange during impulsive encounters.   A similar study was published by Mayer et al.~(2001), who modeled interactions of dIs with the halo of the Milky Way to explore dwarf evolution by tidal stirring.  They do not report on any KDC in their models, which include retrograde interactions.  The two studies differ in that Mayer et al.\ set the satellites on bound orbits and model the larger galaxy as an external potential, thus neglecting dynamical friction.  Our models describe fast, hyperbolic, strongly penetrating encounters with a live primary galaxy.  Our stronger interactions  provide more stringent tests on the formation of counterrotating cores in fly-by interactions.  

In our paper, we first review analytical predictions (\S~\ref{teo}), then describe the models (\S~\ref{models}).  Our results (\S~\ref{results}) will show that, while analytical predictions of angular momentum exchange are accurate, the effects on the rotation curves are small: the dwarf material that absorbs most of the angular momentum absorbs energy as well, and escapes from the dwarf.  Envelope counter-spinning is attained for nearly head-on collisions only, and only if the dwarf does not have a dark matter halo.  
%We conclude that fly-by interactions leading to counterrotation at $R\leq 1.5$\Re\ are exceedingly rare events.  
%\mbremark{"
We conclude that fly-by interactions leading to counterrotation at $R\leq 1.5$\Re\ may have occurred in about 1\% of the dwarf galaxies in a Virgo-type cluster.%}

\begin{figure}
%\centering
\includegraphics[width=8cm]{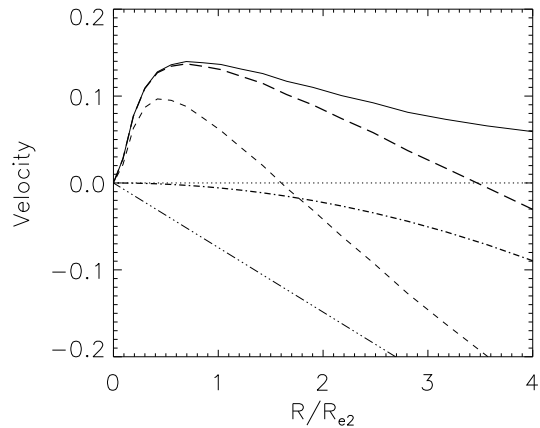}
\caption{Velocity change according to the impulse approximation (eqs. (\ref{eqn:Deltav}) and (\ref{eqn:Deltav2})) for a dwarf galaxy flying through a normal galaxy with mass ratio 10:1 on a parabolic encounter with pericenter distance close to the half-mass radius of the large system (model $CeP$ in Table~\ref{tabfb}). Solid light curve is the initial velocity curve, the dotted-dashed light curve is the velocity change according to eq.~\ref{eqn:Deltav}, the dotted-dashed heavy curve is the velocity change according to eq.~\ref{eqn:Deltav2}, and the light dashed curve and solid dashed curve are the final velocity curves accordingly. \label{fig:Deltav}}
\end{figure}

\section{Theoretical expectations}
 \label{teo}
 
We here review analytical predictions for the exchange of angular momentum in a fly-by encounter of a dwarf galaxy with a giant galaxy.  Our goal is to see under what conditions  the rotation of a dwarf envelope can reverse sign as a result of the fly-by.  
de Rijcke et al.\ (2004) made one such prediction based on the tidal transfer calculation of Som Sunder, Kochhar \& Alladin (1990), yielding a global value for the change of the rotation of the dwarf in terms of its inertia tensor.  We slightly modify that calculation in order to estimate the velocity change as a function of radius in the dwarf, so that diagnostics on counterrotation can be inferred.  

\begin{figure*}
%\centering
\includegraphics[width=18cm]{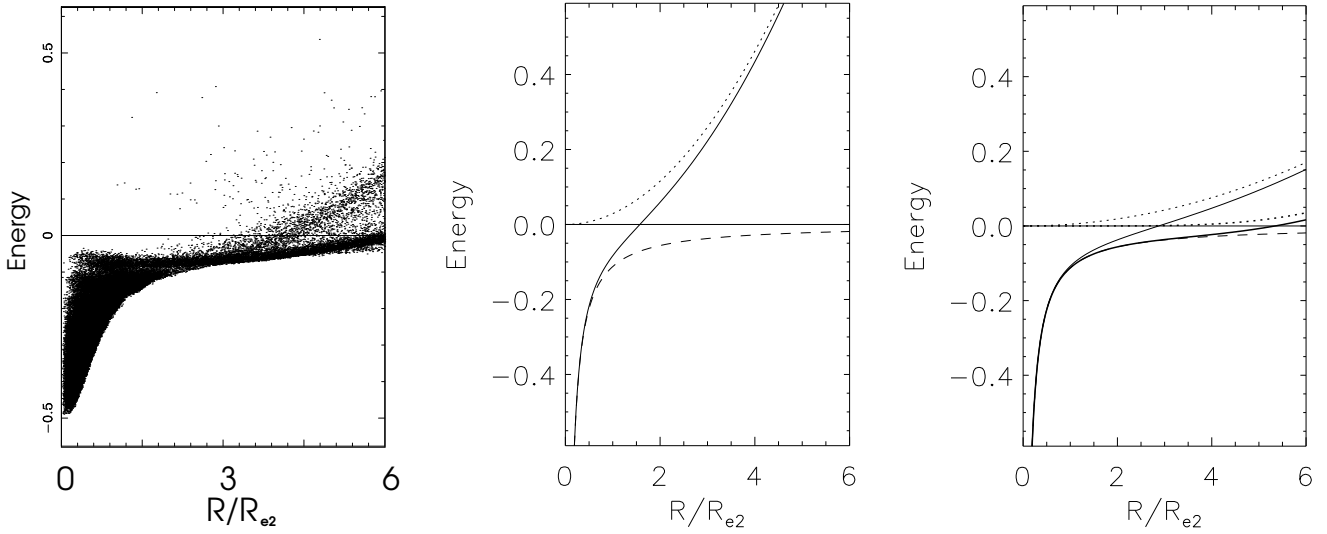}%{deltae_high2.eps}
\caption{Left: Binding energy for the particles of the dwarf galaxy in model $CaHH$ at the end of the fly-by interactions. All particles from the dwarf are plotted. Middle: Energy change according to the impulse approximation (eq.~\ref{eqn:Deltae1}). The dashed curve is the initial binding energy for particles on circular orbits. The dotted line is the expected energy gained according to eq.~\ref{eqn:Deltae1}, and the solid line is the final binding energy. Right: Energy change according to eq.~\ref{eqn:Deltae}, thin lines provide the correction due to the extended nature of the giant, thick lines give the full approximation. Dashed, dotted and solid lines are as before.}\label{fig:deltae}
\end{figure*}

We use the impulse approximation: we assume that the galaxies have no time to reorganize during the encounter, hence the energy injection is all kinetic; reorganization occurs after the encounter has finished.  
We further make use of formulae based on the tidal approximation (Binney \& Tremaine 1987, \S\,7.2(d)), in which the tidal field has been expanded to first order.  
Strictly speaking, the impulse approximation applies only if the encounter time
is much shorter than the internal crossing time,  
and the tidal approximation applies when the impact parameter is much larger than a typical radii of the galaxies.  
While such conditions do not strictly apply to some of the models we simulate, these approximations provide useful estimates of the energy and angular momentum exchange during fast encounters (e.g., Aguilar \& White 1985).  

In the tidal approximation, the velocity change $\Delta\,V_2$ in the stars of dwarf galaxy after a fast interaction with a perturber of mass $M_1$ scales with the dwarf galactocentric radius $R$ as 

\begin{equation}
\mid \Delta {\bf V}_2(R)\mid \sim 2 \frac{G M_{1}}{b^{2}V} R %(1-q_{2}^{2}).
\label{eqn:Deltav}
\end{equation}
\noindent where $b$ the impact parameter of the encounter, $V$ the relative velocity of the galaxies and $G$ is the gravitational constant.  
We have omitted terms of order 1 that depend on the dwarf's axis ratios.  
$\Delta\,V_2(R)$ linearly increases with $R$, hence the perturbation increases in the outer parts.  

In the case when the spin of the dwarf galaxy is opposite to the orbital angular momentum, the change in rotation velocity given by eqn.~\ref{eqn:Deltav} can lead to core-envelope counterspinning.  As an application of eqn.~\ref{eqn:Deltav}, we show  in Figure~\ref{fig:Deltav} the change in the rotation curve expected in a dwarf galaxy as a function of the galactocentric dwarf distance scaled with the effective radius of the dwarf \Re$_2$. The change after a retrograde, parabolic encounter with a 10 times more massive galaxy is shown. The initial rotation curve and effective radius of the dwarf are those used in the $N$-body models, and the pericenter distance is equal to the effective radius of the massive galaxy  (see \S~\ref{models}).  We find that a velocity change as given by eqn.~\ref{eqn:Deltav} leads to a counterrotation radius \Rcr\ $\approx$ 1.5\Re$_2$.  

The tidal approximation used in Equation~\ref{eqn:Deltav} considers the large system as a point mass. We can generalize eqn.~\ref{eqn:Deltav} to take into account the extended nature of the large system. Following Gnedin, Hernquist \& Ostriker (1999) we have:

\begin{equation}
\mid \Delta {\bf V}_2(R)\mid \sim 2 \frac{G M_{1}}{b^{2}V} \frac{R_{peri}}{R_{max}}R^2 %(1-q_{2}^{2}).
\label{eqn:Deltav2}
\end{equation}
\noindent where $R_{peri}$ is the pericenter distance and $R_{max}$ is the cut-off imposed in our initial model for the large galaxy (see \S~\ref{models}). The curves predicted by this new formula can be seen in Figure~\ref{fig:Deltav}, as thick lines. Now the velocity change is smaller and counterrotation is found at larger radii.

Equations \ref{eqn:Deltav} and \ref{eqn:Deltav2} provide $\Delta V_2$ under the assumption that all the transferred angular momentum is kept within the dwarf.  But stars in the dwarf absorb energy as well as angular momentum.
We can estimate the variation of internal energy per unit mass in the secondary at radius $R$ by $\Delta E_2 \sim 0.5 (\Delta V_2)^2$, given that, in the impulse approximation, all the energy absorbed is kinetic:
 \begin{equation}
\Delta E_2 (R) \sim  \frac{G^2 M^2_{1}}{b^{4}V^2} R^2 
\label{eqn:Deltae1}
\end{equation}
\begin{equation}
\Delta E_2 (R) \sim  \frac{G^2 M^2_{1}}{b^{4}V^2} R^2 \big[ \big(\frac{R_{peri}}{R_{max}}\big)^2 (1-\omega\tau)^{-2.5}\big]
\label{eqn:Deltae}
\end{equation}
\noindent where equation~\ref{eqn:Deltae1} comes from Equation~\ref{eqn:Deltav} and eq.~\ref{eqn:Deltae} includes the correction due to the extended nature of our giant galaxy and the term due to the adiabatic corrections, i.e. the corrections that we have to take into account due to the fact that the stars will move during the interaction. Here $\omega$ is the orbital frequency of stars in the satellite and $\tau$ the length of the tidal shock. The exponent -2.5 in fact changes accordingly to the duration of the tidal impulse relative to the dynamical time of the satellite from $2.5-3$ to 1.5 (see Gnedin et al. 1999)

Particles that become unbound as a result of the energy injection carry out part of the angular momentum delivered by the interaction.  The escape of particles is confirmed by Figure~\ref{fig:deltae} which shows the energy budget for one of our models and the expressions given in eq.~\ref{eqn:Deltae1} and~\ref{eqn:Deltae}. Figure~\ref{fig:deltae} (left), shows the distribution of binding energy in the dwarf after the interaction, for the $N$-body realization of model {\it CaHH}.  A tail of unbound particles is seen at $R$/\Re$>3-4$ in this snapshot which correspond to the last time step of our model, when the system has evolved after the encounter. Middle panel shows the effect of the pure tidal approximation (eq.~\ref{eqn:Deltae1}). The figure plots the binding energy per unit mass as a function of radius before the interaction (circular orbits; dashed line); the energy change $\Delta E_2(R)$ from eqn.~\ref{eqn:Deltae1} (dotted line); and the expected final distribution of binding energy (solid line).  Many particles outside 2~\Re\ become unbound and escape the dwarf.  
Failing to account for the angular momentum carried away by these particles should lead to a large overestimate of the angular momentum delivered to the dwarf, and to a corresponding overestimate of the effects on the dwarf rotation curve.
Figure~\ref{fig:deltae} right panel shows the results for the extended nature of the primary (light curves) and for the full eq.~\ref{eqn:Deltae} (heavy curves) which are in closer agreement with the results from our experiments, as we discuss later.

$N$-body fly-by experiments are the ideal tools to determine the true effects of the interactions on the rotation curves of dwarf galaxies.  We address those in the subsequent sections.  

\section{The Models}
\label{models}

Our numerical models simulate the fly-by of a dwarf galaxy about a massive elliptical.  
A King (1966) model was adopted for the dwarf.  The King model has an exponential surface density profile over a large radial range, making it a fair approximation for dE galaxies (Barazza, Binggeli, \& Prugniel 2001).  The model was built using the GalactICS code (Kuijken \& Dubinski 1995). The central
density was 14.45 and the central velocity dispersion
0.714 (in internal units, see below). The model was then scaled down, using $M\sim r^2$ (Fish 1966), to mass 0.1, and was allowed to relax in isolation for fifty half-mass crossing times. During the initial 8-10 crossing times the system expanded, probably due to the softening (about 1/5 of the half-mass radius), then remained in equilibrium for the rest of the run. We used the final relaxed model after the 50 crossing-times as input for our interaction experiments. After scaling and relaxation, the central density and velocity dispersion are 10.40 and 0.302, respectively.  The effective radius is $R_{\rm e2} = 0.178$, while the King concentration index is $C = 0.72$, and the radius that encloses the $99 \%$ of the mass is 2.76. 
We imposed a small rotation by reversing the velocities of 80 \% of the particles chosen randomly, yielding $V_{\rm max}/\sigma_0 = 0.49$.

\input{2670tab}

\begin{figure}
%\centering
\includegraphics[width=8cm,height=5.0cm]{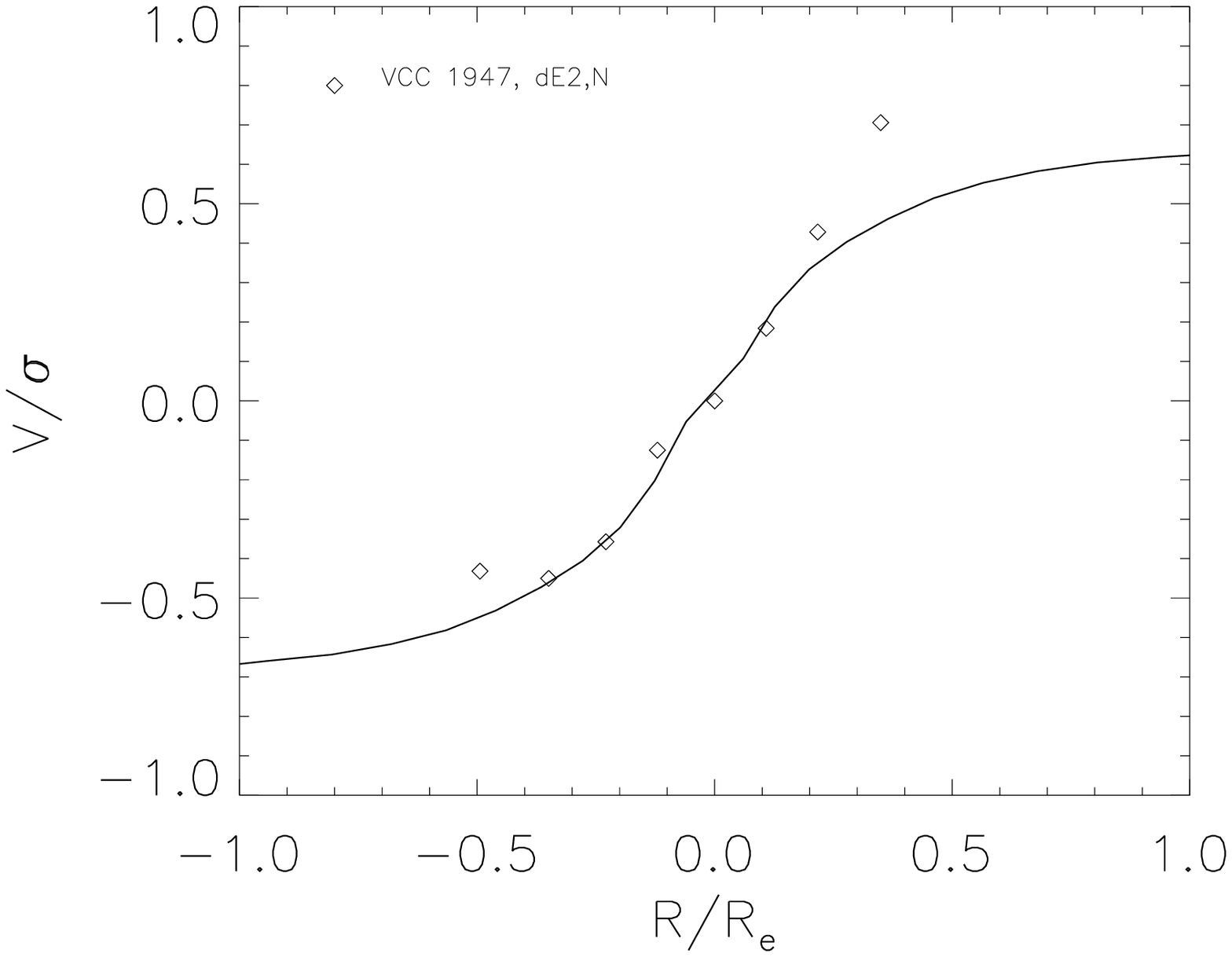}
\vspace{0.cm}
\includegraphics[width=8cm,height=5.0cm]{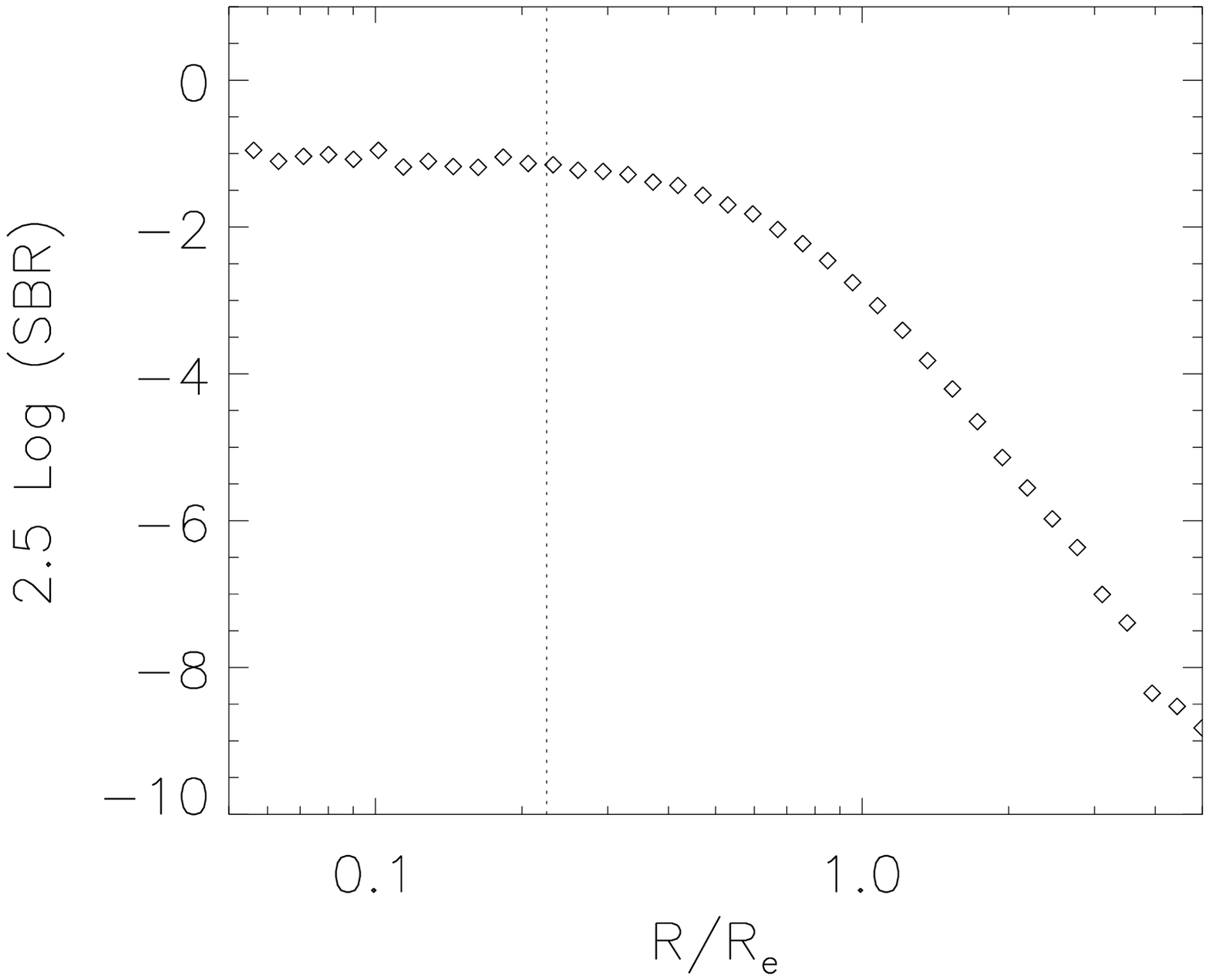}
\caption{Top: rotational support, expressed as $V(R)/\sigma(R)$ vs. $R/R_{\rm e}$ of the initial model (solid line) compared with the dE VCC 1947 (diamonds). Bottom: Surface brightness profile of the initial model.\label{fig:vsini}}
\end{figure}

In order to compare our initial dwarf galaxy model with observed dwarf galaxies we have used the sample by Geha, Guhathakurta \& van der Marel (2003). We use the rotating dE VCC 1947 from their sample as our fiducial initial system. This dE has an absolute magnitude $M_V=-16.32$ and an effective radius of $0.62 \rm{kpc}$. According to this number the mean density inside the effective radius would be $\sim 2 M_{\odot}/pc^3$ (assuming a mean value of $M/L=18M_{\odot}/L_{\odot}$ derived from dwarfs in the Local Group (Mateo 1998)). Using the units given below, our initial system yield a mean density inside the effective radius $1.5 M_{\odot}/pc^3$. Figure~\ref{fig:vsini} top panel compares the rotational support, expressed as $V(R)/\sigma(R)$, for our initial model with the dE VCC 1947. The two systems compare quite well for the observed part. Figure\ref{fig:vsini} bottom panel shows the surface brightness profile of the initial model which can be fitted to a Sersic profile with index $n= 1.07$, similar to dwarf galaxies in the Coma cluster (Aguerri et al.~2005).

The scaling used ($M\sim r^2$) implies a constant surface density for all our dwarf models. This is at odds with the relation between surface brightness and magnitude observed for dwarf galaxies (Ferguson \& Binggeli 1994). However this relation and the one found by Aguerri et al. (2005; see their Fig. 8) show a large scatter that allow us to consider a constant central surface brightness for several magnitudes as a first approximation. Our initial conditions cover 4 magnitudes in absolute magnitude, so for our smaller model $CeP_4$ we should either consider a different M/L to comply with such relation or build the initial model using a different scaling. For consistency reasons we choose the first option, a discussion on the effects of the second one is given below.

In order to test the effects of a dark matter halo around the dwarf, we also run models with a dwarf consisting of a King model inside a halo.  The halo is an Evans model (Kuijken \& Dubinski 1994);  the global central potential is $-4.6$, and the halo asymptotic circular velocity is 2. These parameters give a total mass-to-luminous mass ratio of $M/L\approx 8$, which is on the low regime of values for the dwarf galaxies in the local group (Mateo 1998).

The more massive
galaxy is simulated as an isotropic, non-rotating Jaffe (1983) model. The theoretical half mass radius $r_{\rm J}$ and the total mass $M_{1}$ were set to unity, so that we have a mass ratio for the encounter of 10:1. The initial system was let to evolve in isolation for 10 crossing times. After a mild re-organization, the system reaches equilibrium. Because a cut-off radius was imposed at $r= 10 \times r_{\rm J}$, the true half-mass radius $r_{1/2}$ is 0.83.

Non-dimensional units are adopted throughout, with the constant of
gravity G=1. The theoretical half mass radius of the Jaffe model
$r_{\rm J}$ and the total mass of the elliptical galaxy are also set to unity.  A set of physical units to compare our models with real galaxies would be the following:

\begin{equation}
          [M] = M_{1} = 4\times10^{11} \; {M_{\odot}}, \\
\end{equation}
\begin{equation}
      [L] = r_{\rm J} = 10 \;\rm{kpc}  ,\\
\end{equation}
\begin{equation}
      [T] =  2.4\times10^{7} \;\rm{yr}. \\
\end{equation}
With these, the velocity unit is:

\begin{equation}
      [v] =  414\; \rm{km/s}. \\
\end{equation}
With these units the initial maximum rotation velocity of the dwarf galaxy is
$\approx 60$ km/s.

Initial conditions for the fly-by experiments are listed in Table~\ref{tabfb}. 
Column (1) gives the model code name.  Column (2) lists the mass ratio.  Column (3) gives the spin orientation of the smaller galaxy.  Columns (4), (5), (6) and (7) give the initial separation, orbital eccentricity, pericenter distance and pericenter velocity, respectively, of the orbit the galaxies would follow if they were point masses on Keplerian orbits. The initial separation was always equal to 20 length units, or twice the radius enclosing $99\%$ of the mass of the more massive galaxy.  Inspection of Table~\ref{tabfb} shows that most of our models explore the regime of parabolic or hyperbolic orbits with small pericenter distances \rperi.  

Spin orientation is coded by the first upper-case letters in the model names.  $C$ denotes a counterspinning dwarf.  Given that we are searching for counterrotating signatures, most of our models are $C$ models.  However, we did run one model with dwarf rotation aligned with the orbit (denoted with first letter $R$) and one model with partially misaligned dwarf spin (first letters $CR$).  
 The lower-case letter in the model name codes \rperi, which varies from \rperi = 1/4 of the half mass radius of the dwarf galaxy (models $a$) up to a \rperi\ equal to the radius including $90\%$ of the mass of the elliptical (models $f$).  
For each \rperi\  we run hyperbolic ($H$, $e=1.07$) and parabolic ($P$)
orbits.  For model $a$, the $P$ orbit results in a fast
merger, and is not described further here.  For models $a$ and $b$, we additionally run hyperbolic orbits with
$e=1.003$ and $e=1.043$, which have a longer interaction time (labeled with $HH$). In general the time it takes to the dwarf to cross the inner regions of the giant is less than 1 half-mass crossing time of the giant.

Models listed in the lower part of Table~\ref{tabfb} are variations on the models listed above, in which we selectively change specific model or orbital parameters.  These models are intended to provide clues on the effects that a given parameter may have on the generation of core-envelope counterspinning.  
Model $CeP_G$ is a re-edition of model $CeP$ using five times more particles;  important differences between $CeP$ and $CeP_G$ would point to resolution issues in our results. Model $CeP_\epsilon$ again re-edits model $CeP$ using a smaller value of the softening parameter in the gravitational calculation (see below). 
The effect of a dark matter halo surrounding the dwarf galaxy was investigated with model $CeP_{H}$ which again has identical orbital conditions to $CeP$.
In models $ReP$ and $CReP$, the spin arrangement of model $CeP$ is modified to test the effect of a partial alignment of the secondary spin with the orbit.  
Three models were run with different mass ratio and similar orbital conditions to $CeP$ (mass ratio 5:1, 20:1 and 50:1 for models $CeP2$, $CeP3$ and $CeP4$ respectively).

We also run three more models to study the influence of the initial rotation curve of the dwarf galaxy in the final results. The initial dwarf galaxy for these models was the final state of the dwarf from model $ReP$. We have placed it on an parabolic orbit with anti-parallel spin and $r_{\rm peri}=0.83, 0.2$ and 0.1. These models are named $CeP_{R}$, $CcP_{R}$, and  $CbP_{R}$, respectively.

We used the tree-code {\small GADGET1.1} (Springel et al. 2001) to run our experiments on a Beowulf cluster making use of 16 CPU's. We used $100000$ particles for the King model and $102400$ particles for the Jaffe model. Individual softening for each component was used with $\varepsilon \sim 1/5$ of the half mass radius system ($\varepsilon = 0.04$ for the dwarf and $\varepsilon = 0.1$ for the elliptical; the check-up run $CeP_\epsilon$ uses values of $\varepsilon = 0.015$ and $\varepsilon = 0.045$ respectively). The tolerance parameter was set to $\theta = 0.8$ and quadrupole terms were included in the force calculation. A typical run with $2 \times 10^5$ particles took of the order of $1.8 \times 10^4$ seconds for $7 \times 10^3$ steps.  Conservation of energy was good, with errors well below 0.1\%.   

All models were let to evolve well after the first pass through the pericenter, until the distance between the centroids of the two systems was roughly the initial separation. Dynamical friction prevented models $CaHH$ and $CbP$ from even approaching the initial separation; in these cases, a distance close to apocenter after the first pass through the pericenter was taken for analysis.  Rotation curves for the dwarfs were derived by observing the galaxy models from a point of view perpendicular to the orbital angular momentum and also perpendicular to any tidal feature that might be present in the final dwarf galaxy. Variations on the amplitude of the rotation curve are to be expected when viewed from different points of view. The rotation curves were obtained by placing a slit along its projected major axis with a length of at least 2 length units (=20 kpc, $20 \times r_{1/2}$ of the dwarf galaxy) and a width of 0.3. They give mass-weighted line-of-sight velocities.

\begin{figure}
%\centering
\includegraphics[width=7.0cm,height=5.0cm]{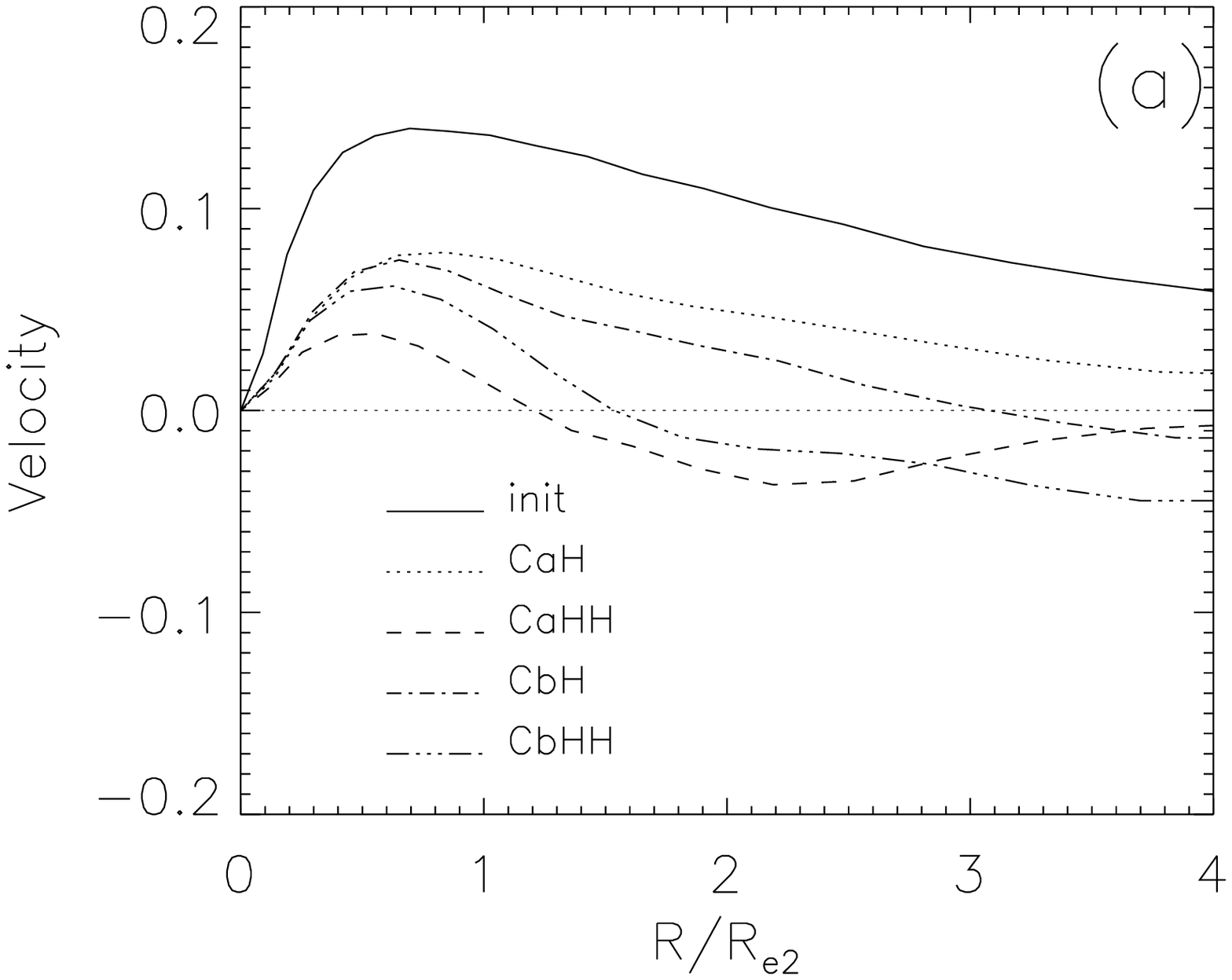}
\vspace{0.cm}
\includegraphics[width=7.0cm,height=5.0cm]{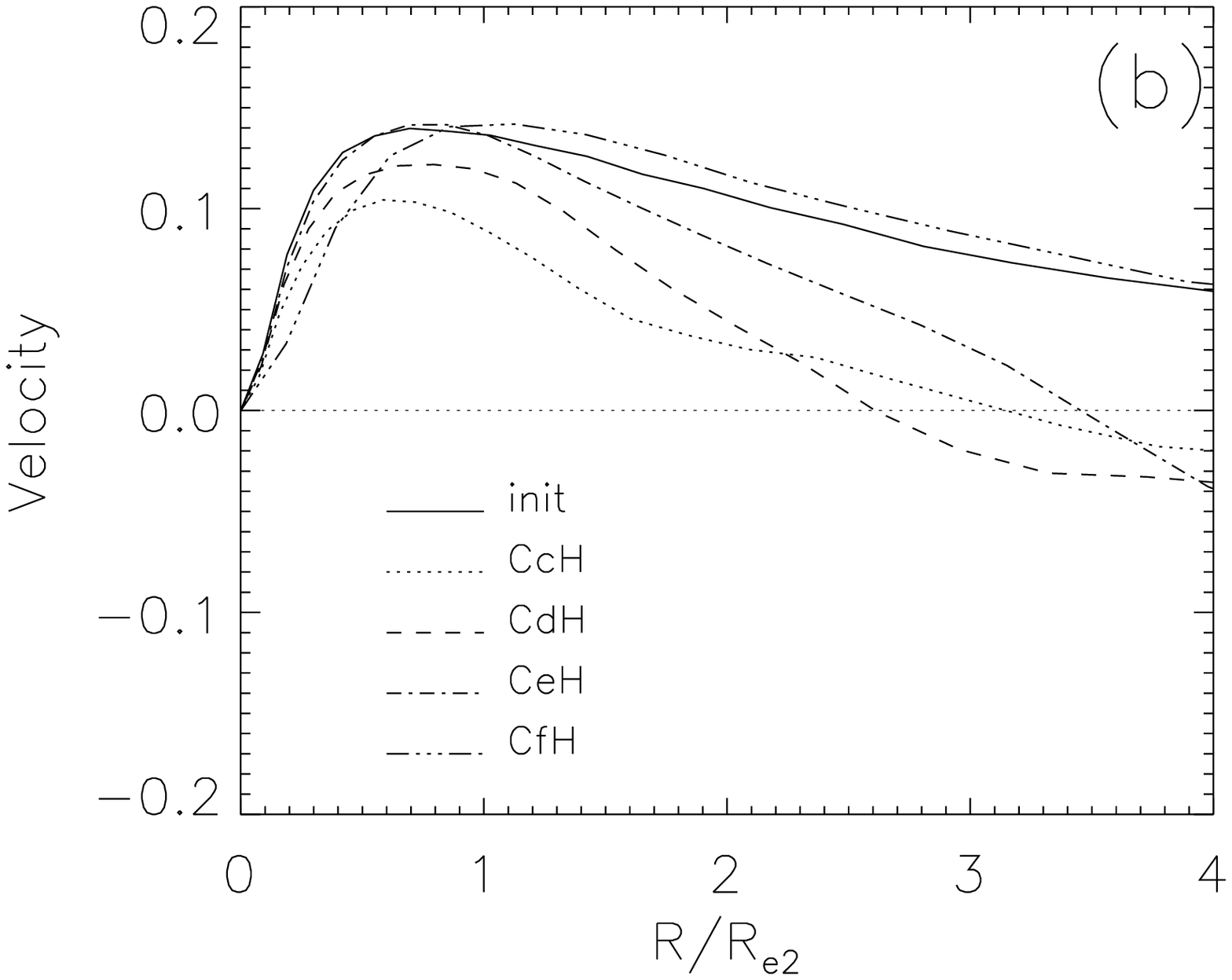}
\vspace{0.cm}
\includegraphics[width=7.0cm,height=5.0cm]{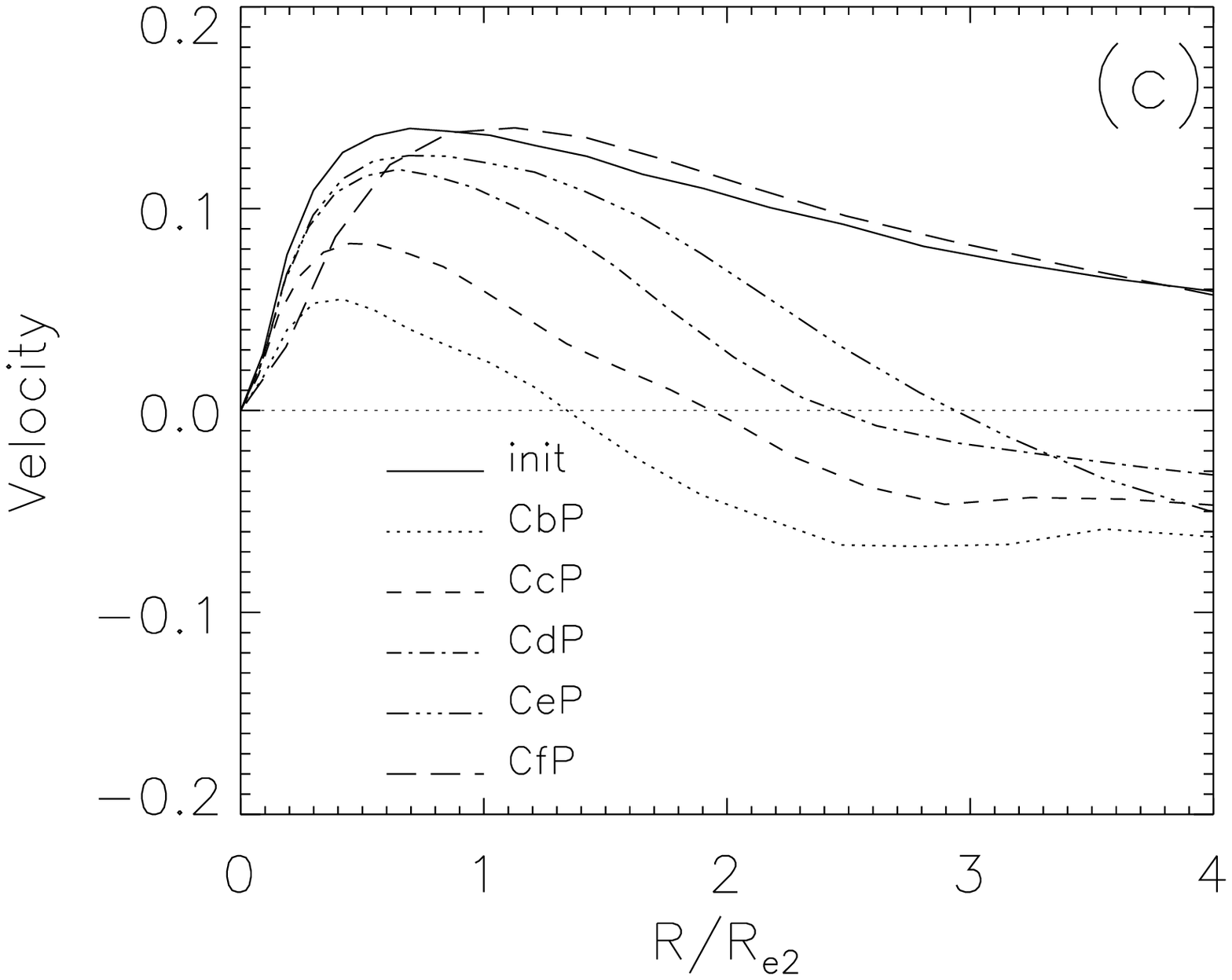}
\caption{Rotation velocity curves for the different models. (a) Models
  with hyperbolic energies and pericenter distance less than \Re\ of
  the dwarf galaxy. (b) Models with hyperbolic energies and pericenter distance larger than \Re\ of the dwarf galaxy. (c) Models with parabolic energies \label{fig:rotc1}}
\end{figure}

\begin{figure}
%\centering
\includegraphics[width=7.0cm,height=5.0cm]{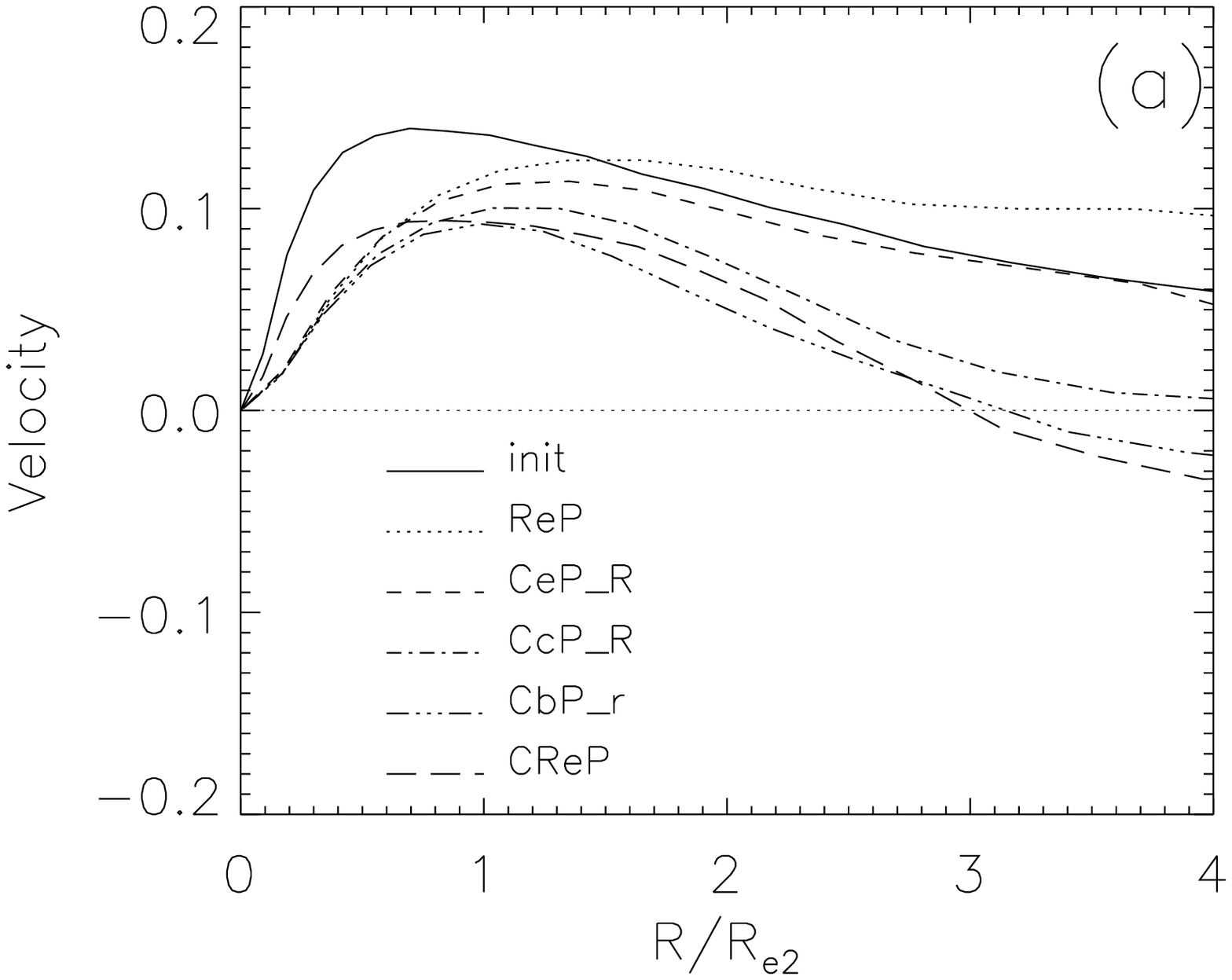}
\vspace{0.cm}
\includegraphics[width=7.0cm,height=5.0cm]{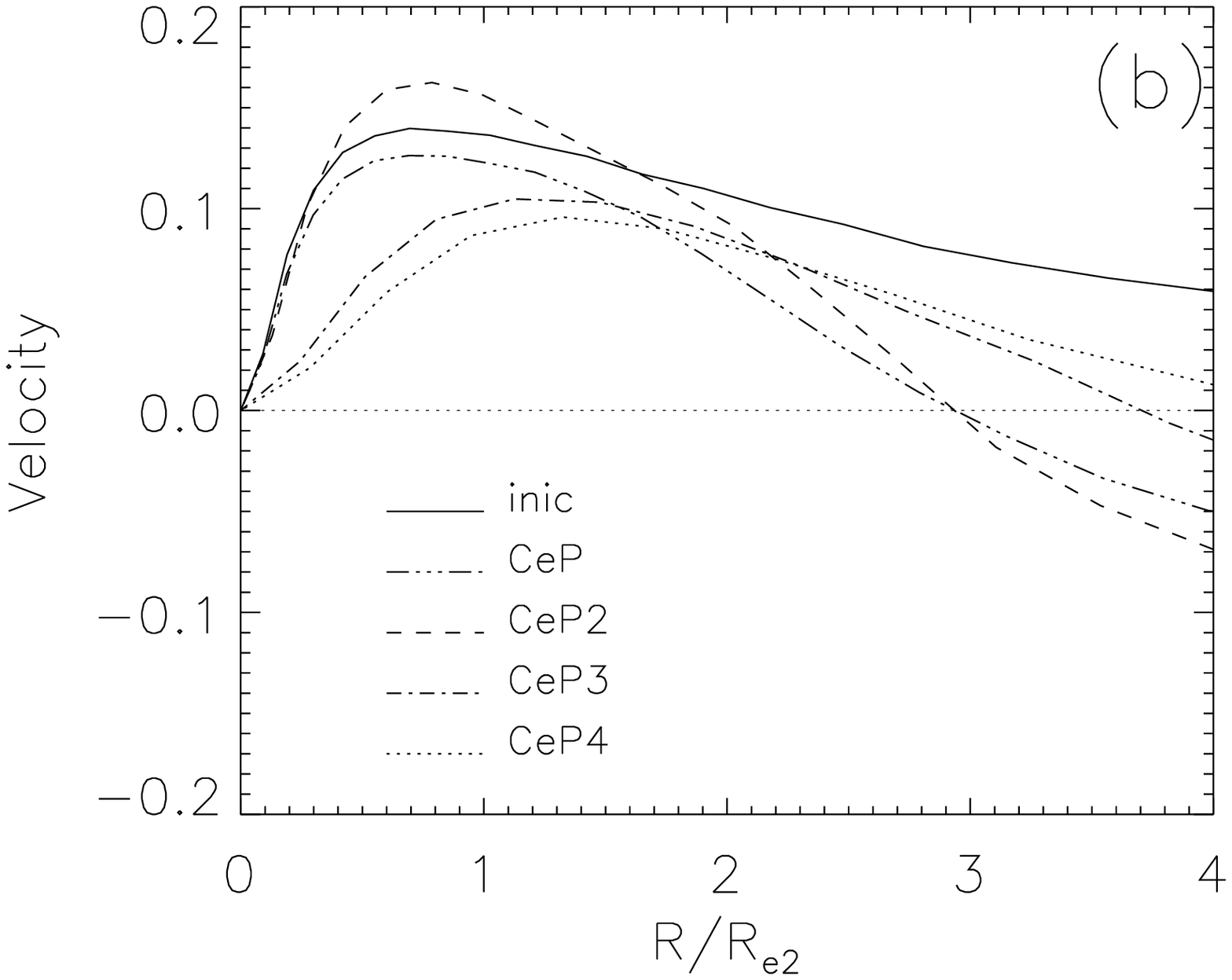}
\vspace{0.cm}
\includegraphics[width=7.0cm,height=5.0cm]{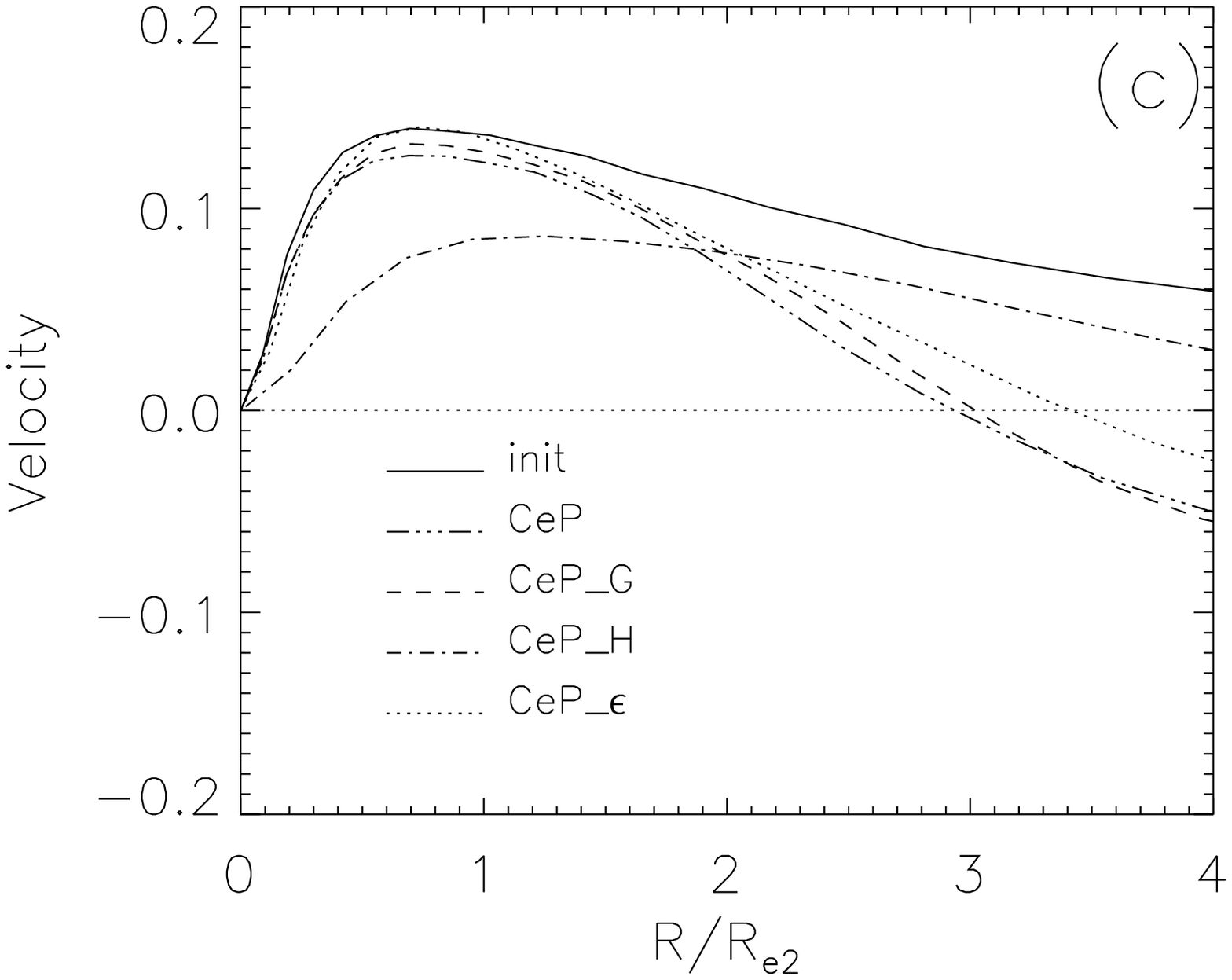}
\caption{Rotation velocity curves for check up models. (a) Models starting from different initial conditions. (b) Models with different initial mass ratio. (c) Model with 5 times more particles, model with a halo and model with a different softening. \label{fig:rotc2}}
\end{figure}

\section{Results} 
\label{results}

Rotation curves for the models from Table 1 are shown in
Figures~\ref{fig:rotc1} and \ref{fig:rotc2}.  Models are grouped in the following way: {\it Fig.~\ref{fig:rotc1}a,b:} hyperbolic encounters, sorted by increasing \rperi.  {\it Fig.~\ref{fig:rotc1}c:} parabolic encounters.  {\it Fig.~\ref{fig:rotc2}a:} variations of the dwarf orientation and rotation curve shape.  {\it Fig.~\ref{fig:rotc2}b:} variations of the mass ratio.  {\it Fig.~\ref{fig:rotc2}c:} changes in the number of particles, and on the presence of a halo in the dwarf.  In each of the panels, the rotation curve for the initial dwarf model is shown for comparison ({\it solid line}).  

Figures~\ref{fig:rotc1} and \ref{fig:rotc2} show that the rotation amplitudes of most models are lower than in the initial model: as expected, the fly-by interactions do produce a transfer of orbital angular momentum to the dwarf.  For retrograde experiments with small \rperi, a counter-rotation (CR) signature appears.  However, a quick inspection reveals that the radius of rotation curve reversal \Rcr\ is significantly larger in the models ($\sim$ 2--3~$R_{\rm e2}$) than in the observed galaxies.  \Rcr\ decreases monotonically as \rperi\ decreases.  In order to have \Rcr\ $\approx R_{\rm e2}$, the encounter has to be almost head-on and mildly energetic (runs $CbP$ and $CaHH$).  

Figure{\it ~\ref{fig:rotc1}a} focuses on the interesting regime of hyperbolic, nearly-head on collisions.   Two values of orbit eccentricity and two values of \rperi\ are shown.  For the two most energetic models ($CaH$ and $CbH$) we find \Rcr/$R_{\rm e2} > 3$, showing that nearly-head on orbits do not ensure that a CR signature appears at radial distances similar to those of real dwarfs.  Angular momentum transfer to the dwarf has low efficiency because the interaction time in these fast encounters is short.  A more realistic CR radius ($R/R_{\rm e2} \approx 1.5$) is obtained in the models with lower eccentricity ($CaHH$ and $CbHH$).  Hence, the parameter space or orbits leading to CR signatures in dwarfs is constrained both in \rperi\ and in orbital energy: fast orbits yield too short an interaction time, while orbits approaching parabolic conditions lead to a fast merger.

We have checked for the stability of the counterrotation by letting the remnant of model $CaHH$ to run in isolation for 50 time units (equivalent to 12 Gyrs). The counterrotation is stable throughout this time.

\section{Dependence of the results on the encounter parameters}
\label{checks}

The results from the previous section provide measurements on how impact parameter and encounter speed determine the onset of envelope CR in fly-by interactions.  Despite the limited number of models we have run, the results give clues that only very strong interactions are capable of imprinting CR at radii similar to those observed in the galaxies FS373 and FS76 (\S~\ref{sec:Introduction}).  We now explore to what degree the results change when other interaction characteristics vary.  It is not our purpose to do a full exploration of parameter space; we just replicate some of the models discussed in \S~\ref{results} varying one single parameter in turn.   We have
changed: spin orientation, initial rotation curve of the dwarf
galaxy, mass-ratio, presence of a dark matter halo in the dwarf galaxy, particle number and softening.  Initial conditions for these models are listed in Table~\ref{tabfb}.  We discuss these models in the following subsections.

\subsection{The spin orientation}
\label{sec:SpinOrientation}

Three of our models have differing spin orientations, while keeping all other parameters unchanged: $CeP$ (dwarf spin antiparallel to the orbit), $ReP$ (spin aligned with the orbit) and $CReP$ (spin at 135$\degr$ to the orbit).  When the dwarf spin is aligned with the orbit ($ReP$, Fig:~\ref{fig:rotc2}a), the external parts of the
dwarf gain angular momentum and rotate faster than
initially. The final rotation curve is almost
flat. No counter-rotating features appear in this model, as expected from theoretical expectations.

The final system for model $CReP$ is quite similar to the model $CeP$, 
with \Rcr/$R_{\rm e2} \approx 3$ (compare with Figs.~\ref{fig:rotc1}c and \ref{fig:rotc2}a). This result suggests that the onset of CR does not depend on the dwarf spin being precisely antiparallel to the orbit.

\subsection{Shape of the dwarf rotation curve}
\label{sec:RotCurveShape}

As a result of the method employed to impart rotation (\S~\ref{models}), our initial dwarf galaxy model has an  outwardly declining rotation curve.  This is in contrast to rotation curves of dwarf galaxies, which typically increase with radius (e.g., Swaters et al.\ 2003; van Zee et al.\ 2004).  
As a simple check on the effects of this difference on our results, we have run experiments using as initial conditions
the dwarf galaxy after the interaction in model $ReP$.  As shown in the previous subsection, the rotation curve after fly-by $ReP$ is shallower in the inner parts than our canonical initial model, and is nearly flat in the outer parts (see Fig.~\ref{fig:rotc2}a).  
We have run 3 models with different \rperi: $CbP_R$, $CcP_R$ and
$CeP_R$. The results are shown in Figure~\ref{fig:rotc2}a.  Again, we find counter-rotation, but \Rcr\ of these new
models is larger than for those of models $CbP$, $CcP$ and $CeP$.  This is
due to the larger rotation in the outer-most region of the new initial model.   
These experiments also show the possible cumulative effect of harassment. After two encounters the CR appears at larger radii than for a single encounter. Thus, in principle, repeated encounters would make even harder to find CR.

\subsection{Models with different mass ratio}
\label{sec:MassRatio}

Models $CeP2$, $CeP3$ and $CeP4$ use the same initial orbital parameters as model $CeP$, but we modify the mass ratio, which is set to 
5:1, 20:1 and 50:1, for models $CeP2$, $CeP3$ and $CeP4$, respectively (Table~\ref{tabfb}). 
The final rotation curves from the three models are shown in  Fig.~\ref{fig:rotc2}b. 
\Rcr/$R_{\rm e2}$\ is quite similar for the two high-mass satellite cases, but is significantly larger for the less massive dwarfs (\Rcr/$R_{\rm e2} \approx 4$). Clearly, smaller mass ratios favor the appearance of counterrotation at smaller radii. We note that \Rcr/$R_{\rm e2}$ increases for larger mass ratios because the effective radius of the smaller system decreases faster than the \Rcr\ does. This is due to the difference in densities introduced by the physical scaling used in our models. Smaller systems are denser in the inner parts and this prevents against the formation of the counter-rotation in the inner parts. Extrapolation of this trend suggests that fly-by interactions might yield smaller values of \Rcr/$R_{\rm e2}$ when applied to the less extreme mass ratios typical of encounters between dwarf galaxies.

\subsection{Presence of dark matter halo in the dwarf galaxy}
\label{sec:Halo}

We have run a model with the same initial orbital parameters as $CeP$ but
including a dark matter halo around the dwarf galaxy (model $CeP_{H}$). Results for this model and model $CeP$ are compared in Fig.~\ref{fig:rotc2}c. We find that model $CeP_H$ shows no counterrotation over the explored radial range.  
Clearly, the dark matter halo of the dwarf galaxy absorbs a large
fraction of the orbital angular momentum leaving no significant
counter-rotating signature on the final system.  The halo-embedded dwarf $CeP_H$ has a shallower inner rotation curve after the interaction, pointing at an enhanced ability to retain orbital angular momentum thanks to the deeper potential well provided by the dark halo.  

\subsection{Models with different number of particles and softening}
\label{sec:LargeN}

In order to check for resolution problems affecting the
results, we have run a model which has the same initial
conditions as $CeP$ but with 5 times more particles (model $CeP{_G}$). The
results are shown in Fig.~\ref{fig:rotc2}c.  We find that the rotation curves of models $CeP$ and $CeP_G$ are identical, suggesting that particle number does not critically influence the results presented here.  

We have also run model $CeP$ with a smaller softening in order to see the effect of having a more accurate force resolution. The results are shown in Fig.~\ref{fig:rotc2}c.  We find that the rotation curves of models $CeP$ and $CeP_\epsilon$ are similar but not identical. 

\begin{figure}[h]
%\centering
\includegraphics[width=8cm]{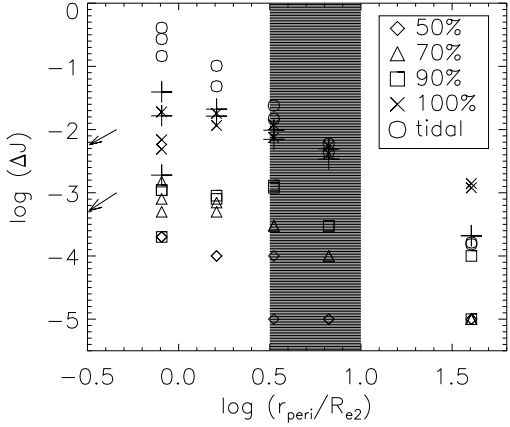}
\vspace{0.cm}
\caption{Angular momentum change versus pericenter distance. The theoretical expectation from the pure impulse approximation for our runs are given as open circles (see eq. \ref{eqn:DeltaJ}) and the full approximation is given as plus signs (see eq. \ref{eqn:DeltaJ2}). The values of the angular momentum change for our models at radii including the  50, 70, 90 and 100$\%$ of the mass of the model are given as diamonds, triangles, squares and crosses respectively. The shadowed area shows the region where the pure tidal approximation applies. The thick and light arrows indicate that $\Delta J$ tends to zero for head-on encounters (\rperi $=0$). \label{fig:DeltaJ}}
\end{figure}

The counterrotation in the new model appears for outer radii than in the fiducial model.

\section{Discussion}
\label{sec:Discussion}

\subsection{Numerical models vs. impulsive approximation}

Comparison of the rotation curves from our $N$-body fly-by interactions (Figs.~\ref{fig:rotc1} and \ref{fig:rotc2}) with the theoretical expectations (\S~\ref{teo}, and Fig.~\ref{fig:Deltav}) shows that the fly-by interactions are far less effective in reversing the velocity sign than expected from the pure impulse approximation. The full treatment must be applied in order to account for the observed behaviour.

A first clue to this behaviour is given by the radial distribution of $\Delta(J_2)$, the angular momentum deposited in the secondary.  In Figure~\ref{fig:DeltaJ} we plot $\Delta(J_2)$ integrated out to radii including 50\%, 70\%, 90\% and 100\% of the secondary mass, against the pericenter distance.  The figure indicates that, for all the models, $\Delta(J_2)$ in the outer 10\% of the mass is over 10 times larger than that in the inner 90\% of the secondary mass.  As suggested by Figure~\ref{fig:deltae}, the outer parts of the satellite become unbound after the interaction.  Therefore, over 90\% of the transferred angular momentum escapes the dwarf in tidal tails.  The effects of the interaction on the dwarf rotation curve must be weaker than predicted by eqn.~\ref{eqn:Deltav}.

A second clue to the discrepancy is provided by comparing the true angular momentum transfer to that predicted by the pure impulse approximation in the tidal limit.  Using eqn.~\ref{eqn:Deltav}, $\Delta J_2(R)  \sim M_{2}R \Delta V_2(R)$, and $I_{22}(R) \sim M_{2} R^2 /3$, we get:
\begin{equation}
\Delta J_2(R) \sim 2 \frac{G M_{1}}{b^{2}V} I_{22}(R), %(1-q_{2}^{2}),
\label{eqn:DeltaJ}
\end{equation}
where $I_{22}$ is the smallest eigenvalue of the
inertia tensor.   %, and $q_{2}$ is the axis ratio of the dwarf galaxy.
$\Delta(J_2)$ computed with eqn.~\ref{eqn:DeltaJ} for the entire secondary is plotted with circles in Figure~\ref{fig:DeltaJ}.  We find that eqn.~\ref{eqn:DeltaJ} accurately predicts the total angular momentum transfer (crosses in Fig.~\ref{fig:DeltaJ}) for experiments with intermediate pericenter distances, but it overestimates $\Delta(J_2)$ at small \rperi, and it underestimates $\Delta(J_2)$ at large \rperi.  The overestimate of $\Delta(J_2)$ at small \rperi\ aggravates the discrepancy between the prediction of the impulse approximation and the fly-by experiments.  As we diminish \rperi, $\Delta(J_2)$ tends to zero, as we approach head-on conditions (indicated with arrows in Figure~\ref{fig:DeltaJ}).  At the other end, $\log ($\rperi$/R_{\rm e2}) \ga 1$, the discrepancy between measured and predicted $\Delta(J_2)$ may be due to the fact that the interaction time is comparable to the dynamical time of the secondary, in contradiction with the assumptions of the impulse approximation.

Figure~\ref{fig:DeltaJ} supports the findings of Aguilar \& White (1985), who show that the tidal approximation gives reasonable results for impact parameters of order $5 \times r_h$, with $r_h$ the median radius of the small system.  This range of impact parameters, shaded in figure~\ref{fig:DeltaJ}, encompasses the models where eqn.~\ref{eqn:DeltaJ} reproduces the measured $\Delta(J_2)$.  

The discrepancies are less extreme if we take into account the full treatment of the impulse approximation, and use eqn.~\ref{eqn:Deltav2}. Then we get:
\begin{equation}
\Delta J_2(R) = 2 \frac{G M_{1}}{b^{2}V} I_{22}(R)  R \big(\frac{R_{peri}}{R_{max}}\big) (1-\omega\tau)^{-2.5}
\label{eqn:DeltaJ2}
\end{equation}
This is given in Figure~\ref{fig:DeltaJ} as plus signs. The agreement is quite good even for the inner parts. For large pericenter distances the approximation does not apply due to the long duration of the tidal impulse and we are still not able to account for the angular momentum carried away by the tidal tails.

\subsection{Comparison with observations}

As described in \S~\ref{sec:Introduction},  de Rijcke et al.~(2004) have found KDCs in two dE elliptical galaxies belonging to the NGC~5044 and NGC~3258 groups. These galaxies show CR features at
\Rcr/\Re~$\approx 1-1.5$, which de Rijcke et al.\ explain as an effect of harassment interactions.   
Other observational studies of dE rotation curves rarely reach the depths required to map rotation to $\sim$\Re, hence determining the frequency of CR in dEs is subject to strong observational biases.  The studies of Virgo dwarfs (Pedraz et al.\ 2002;
van Zee et al.\ 2004) do show a fraction of objects with kinematic substructure, such as asymmetric or otherwise complex rotation curves.   In particular, 1 out of 6 dEs from Pedraz et al.\ (2002) shows hints of envelope counterrotation.  %

\begin{figure}[h]
%\centering
\includegraphics[width=8cm]{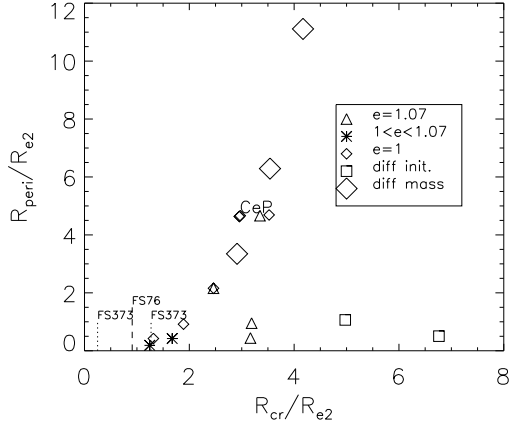}
\vspace{0.cm}
\caption{Pericenter distance vs. CR radii in units of the secondary effective radius. The different symbols indicate the different orbital configurations.\label{fig:rprc}}
\end{figure}

Figure~\ref{fig:rprc} summarizes our results showing the counterrotation radii vs. the pericenter distance in units of the secondary effective radius. The values for the CR in the two dE from de Rijcke et al. (2004) are given as a comparison. Interactions such as our models $a$ and $b$ may explain the
CR features observed in the two dwarf galaxies studied by de Rijcke et al.~(2004).  

However, our models tend to produce CR signatures at larger radii than is observed.  The required small impact parameters are a strong restriction for the types of collisions that may lead to counterrotation.  In the next subsection we estimate the probability of such collisions in a cluster environment.

\subsection{The harassment scenario for dwarf CR}

To determine whether harassment is a likely mechanism for the formation of CR, we estimate here the probability of small impact parameter collisions.  Rather than a group, we choose a cluster environment since it contains large numbers of giant and dwarf galaxies, which allows for a statistical calculation.  We use a very simplified picture of the dynamical state of a cluster, which we describe with a single, representative number density of large galaxies, and we assume that velocities are uncorrelated.  The cross-section $\SigCR$ for collisions leading to core-envelope CR is approximately

\begin{equation}
\SigCR \approx 0.15\,{\rm \pi}\,b^2
\end{equation}
where $b\approx 5$ kpc is a typical impact parameter for experiments $Ca$ and $Cb$, and the numerical factor is the fraction of encounters with spins within 45$^\circ$ from anti-parallel.  The mean-free path $\lambdaCR$ between collisions leading to CR is

\begin{equation}
\lambdaCR = \frac{1}{n\,\SigCR}
\end{equation}
where $n$ is the number density of bright galaxies in the cluster.  For Virgo, taking $n\approx 50$ gal\,Mpc$^{-3}$ (Binggeli et al.~1987), we infer $\lambdaCR\approx 1700$~Mpc.  In a time interval $\Delta t$, a dwarf in a cluster with one-dimensional velocity dispersion $\sig1D$ will travel a distance $L = \sqrt(3)\,\sig1D\,\Delta\,t$, hence the probability of a CR collision may be expressed as 

\begin{equation}
P_{\rm CR} = L / \lambdaCR = \sqrt{3}\,\sig1D\,\Delta t\,n\,\SigCR,
\end{equation}
then, for \Ndwarfs\ dwarfs, the number of them prone to have a CR collision is 

\begin{equation}
N_{\rm CR} = \Ndwarfs\,\sqrt{3}\,\sig1D\,\Delta\,t\,n\,\SigCR,
\end{equation}
which, for values typical of the Virgo cluster (Binggeli et al.~1985), yields 

\begin{eqnarray}
\lefteqn{N_{\rm CR} =} \nonumber \\ 
& & 16 \left(\frac{\Ndwarfs}{2000}\right)\left(\frac{\sig1D}{800\,\kms}\right)
\left(\frac{\Delta t}{10\,\rm{Gyr}}\right)\left(\frac{n}{50\,\rm{Mpc^{-3}}}\right)\left(\frac{b}{5\,\rm{kpc}}\right)^2
\end{eqnarray}

In other words, we may expect to find 8 dwarfs showing CR in a cluster like Virgo. This means that, at face value, about 1\% of dwarfs in a Virgo-type cluster may experience a CR collision in 10~Gyr.  Our estimate is admittedly crude.  Some of the collisions will be too slow, or too fast, for an efficient transfer of angular momentum; dwarfs embedded in their own dark matter halos hardly respond to the interaction; internal velocities in the cluster are not uncorrelated; and, dwarfs in elongated orbits spend most of their lifetimes in the low-density outskirts of the cluster.  On the other hand, the cross-section for CR is bigger in interactions with a smaller mass ratio than modeled here if, as suggested in \S~\ref{sec:MassRatio}, lower mass ratios lead to smaller \Rcr/$R_{\rm e2}$.  

Finally, more complex prescriptions for the structure and internal dynamics of the dwarf may lead to different results on \Rcr.  For instance, a dwarf consisting of a disk embedded in a non-rotating envelope requires less angular momentum exchange to generate a CR signature, as shown by the fly-by models explored by Hau \& Thomson (1994). The scaling used for our initial conditions (Fish 1966) implies a constant  surface density. As said in section~\ref{models},  this could be consistent for a range of magnitudes for the dwarf galaxies studied by Aguerri et al. (2005). A different scaling may reproduce the general trend observed where fainter galaxies have a lower surface brightness. This would imply that the volume density would not increase with decreasing mass as fast as in our scaling, changing somewhat the CR radius for the smaller systems.

Despite these caveats, our conclusions hold and in a cluster like Virgo $\sim$ 1$\%$\ of dwarf galaxies may present counter-rotation formed by harassment.

\clearpage

%% the file, uncomment the \plotone, \plottwo, and
%% If you need a layout that cannot be achieved with \plotone or
%% \plottwo, you can invoke the graphicx package directly with the
%% \includegraphics command or use \plotfiddle. For more information,
%% please see the tutorial on "Using Electronic Art with AASTeX" in the
%% documentation section at the AASTeX Web site,
%% http://www.journals.uchicago.edu/AAS/AASTeX.
%%
%% The examples below also include sample markup for submission of
%% supplemental electronic materials. As always, be sure to check
%% the instructions to authors for the journal you are submitting to
%% for specific submissions guidelines as they vary from

%% This example uses \plotone to include an EPS file scaled to
%% 80% of its natural size with \epsscale. Its caption
%% has been written to indicate that additional figure parts will be
%% available in the electronic journal.

%\begin{figure}
%\epsscale{.80}
%\plotone{f1.eps}
%\caption{Derived spectra for 3C138 \citep[see][]{heiles03}. Plots for all sourc%es are available
%in the electronic edition of {\it The Astrophysical Journal}.\label{fig1}}
%\end{figure}

\clearpage

\clearpage

\clearpage

\end{document}

%% file: 2670tab.tex
\begin{table}
%\begin{center}
\caption{Initial configurations for fly-by models .\label{tabfb}}
\centering
\begin{tabular}{@{}ccccccc}
%{\large
%\label{tabdbh}
%\begin{tabular}{||c|c|c|c|c|c|c|c||}
\hline
\hline
{\bf Mod.} &{\bf $\frac{M_1}{M_2}$}& {\bf $(\theta_2,\phi_2)$} & {\bf $r_i$} & {\bf $e$}& {\rperi}& {\bf $V_{peri,K}$}\\
(1) & (2) & (3) & (4) & (5) & (6) & (7) \\
\hline
$CaH$ & 10:1 & (180,0) & 20 & 1.07 & 0.05 & 6.77 \\ %(2800) \\
$CaHH$ & 10:1 & (180,0) & 20 & 1.003 & 0.05& 6.64 \\ %(2750)\\
$CbH$ & 10:1 & (180,0) & 20 & 1.07 & 0.10& 4.77 \\ %(1975) \\
$CbHH$ & 10:1 & (180,0) & 20 & 1.013 & 0.10& 4.70 \\ %(1945) \\
$CbP$ & 10:1 & (180,0) & 20 & 1.00 & 0.10& 4.69 \\ %(1940) \\
$CcP$ & 10:1 & (180,0) & 20 & 1.00 & 0.20& 3.32 \\ %(1375) \\
$CcH$ & 10:1 & (180,0) & 20 & 1.07 & 0.20& 3.37 \\ %(1395)\\
$CdP$ & 10:1 & (180,0) & 20 & 1.00 & 0.41& 2.31 \\ %(956)\\
$CdH$ & 10:1 & (180,0) & 20 & 1.07 & 0.41& 2.35 \\ %(973)\\
$CeP$ & 10:1 & (180,0) & 20 & 1.00 & 0.83& 1.63 \\ %(675)\\
$CeH$ & 10:1 & (180,0) & 20 & 1.07 & 0.83& 1.66 \\ %(687)\\
$CfP$ & 10:1 & (180,0) & 20 & 1.00 & 5.00& 0.66 \\ %(275)\\
$CfH$ & 10:1 & (180,0) & 20 & 1.07 & 5.00& 0.68 \\ %(280)\\  
\hline
$CeP_G$ & 10:1 & (180,0) & 20 & 1.00 & 0.83& 1.63 \\ %(675) \\
$CeP_\epsilon$ & 10:1 & (180,0) & 20 & 1.00 & 0.83& 1.63 \\ %(675) \\
$CeP_H$ & 10:1 & (180,0) & 20 & 1.00 & 0.83& 1.63 \\ %(675)\\
$CeP_R$ & 10:1 & (180,0) & 20 & 1.00 & 0.83& 1.63 \\ %(675)\\
$CcP_R$ & 10:1 & (180,0) & 20 & 1.00 & 0.2& 3.32 \\ %(1375)\\
$CbP_R$ & 10:1 & (180,0) & 20 & 1.00 & 0.1& 4.69 \\ %(1940) \\
%%%%%\hline
$ReP$ & 10:1 & (0,0) & 20 & 1.00 & 0.83& 1.63 \\ %(675) \\
%\hline
$CReP$ & 10:1 & (135,0) & 20 & 1.00 & 0.83& 1.63 \\ %(675) \\
%\hline  
$CeP2$ & 5:1 & (180,0) & 20 & 1.00 & 0.83& 1.71 \\ %(707) \\
$CeP3$ & 20:1 & (180,0) & 20 & 1.00 & 0.83& 1.60 \\ %(660)\\
$CeP4$ & 50:1 & (180,0) & 20 & 1.00 & 0.83& 1.57 \\ %(650)\\
\hline 

\end{tabular}
%}
%\end{center}
%\end{minipage}
\end{table}
%\end{longtable}

%% file: 2670main.bbl
\begin{thebibliography}{}

\bibitem[Aguerri, Iglesias-Paramo, Vilchez, \& Mu{\~ n}oz-Tu{\~ n}{\' 
o}n(2004)]{} Aguerri, J.~A.~L., Iglesias-Paramo, J., 
Vilchez, J.~M., \& Mu{\~ n}oz-Tu{\~ n}{\' o}n, C.~2004, \aj, 127,
1344
\bibitem[Aguerri, Iglesias-Paramo, Vilchez, \& Mu{\~ n}oz-Tu{\~ n}{\' 
o}n(2005)]{} Aguerri, J.~A.~L., Iglesias-Paramo, J., 
Vilchez, J.~M., Mu{\~ n}oz-Tu{\~ n}{\' o}n, C. \& Sanchez-Janssen, R.,~2005, in press
\bibitem[Aguerri, Balcells, \& Peletier(2001)]{} 
Aguerri, J.~A.~L., Balcells, M., \& Peletier, R.~F.~2001, \aap, 367, 428

\bibitem[Aguilar \& White(1985)]{1985ApJ...295..374A} Aguilar, L.~A., \& 
White, S.~D.~M.\ 1985, \apj, 295, 374 

\bibitem[Balcells \& Gonz{\' a}lez(1998)]{} Balcells, M., \& Gonz{\'a}lez, A.~C.~1998, \apjl, 505, L109

\bibitem[Balcells \& Quinn(1990)]{} Balcells, M., \& 
Quinn, P.~J.~1990, \apj, 361, 381 

\bibitem[Barazza, Binggeli, \& Prugniel(2001)]{} 
Barazza, F.~D., Binggeli, B., \& Prugniel P.~2001, \aap, 373, 12 

\bibitem[Bender (1990)]{} Bender, R.~1990, \aap, 229, 441

\bibitem[Binggeli, Sandage, \& Tammann(1985)]{} 
Binggeli, B., Sandage, A., \& Tammann, G.~A.~1985, \aj, 90, 1681

\bibitem[Binggeli, Sandage, \& Tammann(1988)]{} 
Binggeli, B., Sandage, A., \& Tammann, G.~A.~1988, \araa, 26, 509

\bibitem[Binggeli, Tammann \& Sandage(1987)]{} 
Binggeli, B., Tammann, G.~A., \& Sandage, A.~1987, \apj, 94, 251

\bibitem[]{} Binney, J., \& Tremaine, S.~1987, Galactic Dynamics, Princeton University Press, Princeton

%\bibitem[Cappellari et al.(2002)]{} Cappellari, M., 
%Verolme, E.~K., van der Marel, R.~P., Kleijn, G.~A.~V., Illingworth, G.~D., 
%Franx, M., Carollo, C.~M., \& de Zeeuw, P.~T.\ 2002, \apj, 578, 787

%\bibitem[]{} Corsini et al.

%\bibitem[Davies \& Hunter(1997)]{} Davies, C.~L.~\& Hunter, J.~H.\ 1997, \apj, 484, 79

\bibitem[Dekel \& Silk(1986)]{} Dekel, A., \& Silk, J.~1986, \apj,
  303, 39 

\bibitem[De Rijcke \& Debattista(2004)]{} De Rijcke, S., \& Debattista,
  V.~P.~2004, \apjl, 603, L25 

%\bibitem[de Vaucouleurs(1954)]{} de Vaucouleurs, G.\ 1954, The Observatory, 74, 23

\bibitem[De Young \& Gallagher(1990)]{} De Young, D.~S., \& Gallagher,
 J.~S.~1990, \apjl, 356, L15

%\bibitem[Efstathiou, Ellis, \& Carter(1982)]{} Efstathiou, G., Ellis, R.~S., \& Carter, D.\ 1982, \mnras, 201, 975

\bibitem[Evans \& Collett(1994)]{} Evans N.~W., \& Collett, J.~L.~1994, \apjl, 420, L67

\bibitem[Ferguson \& Binggeli(1994)]{} Ferguson, 
H.~C., \& Binggeli B.~1994, \aapr, 6, 67

\bibitem[Ferguson \& Sandage(1989)]{}Ferguson, H.~C., \& 
Sandage, A.~1989, \apjl, 346, L53

\bibitem[Fish (1964)]{} Fish, R.~A.,~1964, \apj, 139, 284

\bibitem[Gallagher \& Wyse(1994)]{} Gallagher, J.~S., \&
 Wyse, R.~F.~G.~1994, \pasp, 106, 1225

\bibitem[Geha, Guhathakurta, \& van der Marel(2003)]{2003AJ...126.1794} 
Geha, M., Guhathakurta, P., \& van der Marel, R.~P.\ 2003, \aj, 126, 1794

%\bibitem[]{}Gonz\'alez-Garc\'{\i}a, A.~C., \& van Albada, T.~S.~2004, \mnras, submitted
%\bibitem[]{}Gonz\'alez-Garc\'{\i}a, A.~C., \& Balcells, M.~2004, \mnras, submitted
\bibitem[Gendin et al.(1999)]{1999ApJ...514..109} Gendin, O.~Y., 
Hernquist, L.~\& Ostriker, J.~H.\ 1998, \apj, 514, 109 
\bibitem[Harsoula \& Voglis(1998)]{1998A&A...335..431H} Harsoula, M.~\& 
Voglis, N.\ 1998, \aap, 335, 431 

\bibitem[Hau \& Thomson(1994)]{1994MNRAS.270L..23H} Hau, G.~K.~T.~\& 
Thomson, R.~C.\ 1994, \mnras, 270, L23 

%\bibitem[Heisler, Merritt, \& Schwarzschild(1982)]{} Heisler, J.,
%  Merritt, D., \& Schwarzschild, M.\ 1982, \apj, 258, 490 

\bibitem[]{} Hernquist, L., \& Barnes, J.~1991, \nat, 354, 210

\bibitem[Jaffe(1983)]{} Jaffe, W.~1983, \mnras, 202, 995 

\bibitem[King(1966)]{} King, I.~R.~1966, \aj, 71, 64 

\bibitem[Kuijken \& Dubinski (1994)]{} Kuijken, K., \& Dubinski, J.~1994, \mnras, 269, 13

\bibitem[Kuijken \& Dubinski(1995)]{} Kuijken, K., \& Dubinski, J.~1995, \mnras, 277, 1341

\bibitem[Lin \& Faber(1983)]{} Lin, D.~N.~C., \& Faber, S.~M.~1983,
  \apjl, 266, L21

\bibitem[Mateo(1998)]{} Mateo, M.~L.~1998, \araa, 36, 
435

\bibitem[Mayer et al.~(2001)]{} Mayer, L., Governato, F., Colpi, M., Moore B., \& Quinn T.~2001, \apj, 547, L123

\bibitem[Mo, Mao \& White~(1998)]{} Mo, H.~J., Mao, S., \&  White, S. D. M.~1998, \mnras, 295, 319

\bibitem[Moore et al.~(1996)]{} Moore, B., Katz, N., Lake, G., Dressler, A., \& Oemler A.~Jr.~1996, \nat, 379, 613
%\bibitem[]{}Mediavilla et al.~1998.

\bibitem[]{} Oosterloo, T., Balcells, M.~, \& Carter D.~1994, \mnras, 266, L10

\bibitem[Pedraz et al.(2002)]{2002MNRAS.332L..59P} Pedraz, S., Gorgas, J., 
Cardiel, N., S{\' a}nchez-Bl{\' a}zquez, P., \& Guzm{\' a}n, R.~2002, 
\mnras, 332, L59 

%\bibitem[Prada, Gutierrez, Peletier, \& McKeith(1996)]{} Prada, F., Gutierrez, C.~M., Peletier, R.~F., \& McKeith, C.~D.\ 1996, 
%\apjl, 463, L9 

\bibitem[Richer \& McCall(1995)]{} Richer, M.~G.,~\& McCall, M.~L.~1995, \apj, 445, 642

%\bibitem[Rubin(1994)]{} Rubin, V.~C.\ 1994, \aj, 108, 456 

\bibitem[Skillman, Kennicutt, \& Hodge(1989)]{} Skillman, E.~D.,
  Kennicutt, R.~C.,~\& Hodge P.~W.~1989, \apj, 347, 875

%\bibitem[]{} Smulders \& Balcells, 1995

\bibitem[Som Sunder, Kochhar, \& Alladin(1990)]{1990MNRAS.244..424S} Som 
Sunder, G., Kochhar, R.~K., \& Alladin, S.~M.\ 1990, \mnras, 244, 424 

\bibitem[Springel, Yoshida, \& White(2001)]{} Springel,
 V., Yoshida, N.,~\& White S.~D.~M.~2001, New Astronomy, 6, 79

\bibitem[Tremaine \& Yu(2000)]{} Tremaine, S.,~\& Yu, Q.~2000, \mnras, 319, 1 

%\bibitem[van Albada, van Leer, \& Roberts(1982)]{} van Albada, G.~D., van Leer, B., \& Roberts, W.~W.\ 1982, \aap, 108, 76

\bibitem[van Zee, Skillman, \& Haynes(2004)]{} van Zee, L., Skillman, E.~D.,~\& Haynes, M.~P.~2004, \aj, 128, 121

\bibitem[Swaters, Madore, van den Bosch, \& 
Balcells(2003)]{2003ApJ...583..732S} Swaters, R.~A., Madore, B.~F., van den 
Bosch, F.~C.,~\& Balcells, M.~2003, \apj, 583, 732 

\end{thebibliography}
